\documentclass[twocolumn,showpacs,preprintnumbers,amsmath,amssymb,superscriptaddress,prb]{revtex4}

\usepackage{graphicx}
\usepackage{dcolumn}
\usepackage{bm}

\begin{document}

\setlength{\parskip}{0 pt}

\preprint{APS/123-QED}

\bibliographystyle{unsrt}

\title{Non-gapped Fermi surfaces, quasiparticles and the anomalous temperature
dependence of the near-$E_F$ electronic states in the CMR oxide
La$_{2-2x}$Sr$_{1+2x}$Mn$_2$O$_7$ with $x=0.36$}

\author{S. de Jong}
\email{sdejong@science.uva.nl}
\affiliation{Van der Waals-Zeeman Institute, University of Amsterdam, NL-1018XE
Amsterdam, The Netherlands}
\author{Y. Huang}
\affiliation{Van der Waals-Zeeman Institute, University of Amsterdam, NL-1018XE Amsterdam, The Netherlands}
\author{I. Santoso}
\affiliation{Van der Waals-Zeeman Institute, University of Amsterdam, NL-1018XE Amsterdam, The Netherlands}
\author{F. Massee}
\affiliation{Van der Waals-Zeeman Institute, University of Amsterdam, NL-1018XE Amsterdam, The Netherlands}
\author{R. Follath}
\affiliation{BESSY GmbH, Albert-Einstein-Strasse 15, 12489 Berlin, Germany}
\author{O. Schwarzkopf}
\affiliation{BESSY GmbH, Albert-Einstein-Strasse 15, 12489 Berlin, Germany}
\author{L. Patthey}
\affiliation{Paul Scherrer Institute, Swiss Light Source, CH-5232
Villigen, Switzerland}
\author{M. Shi}
\affiliation{Paul Scherrer Institute, Swiss Light Source, CH-5232
Villigen, Switzerland}
\author{M. S. Golden}
\email{mgolden@science.uva.nl}
\affiliation{Van der Waals-Zeeman Institute, University of Amsterdam, NL-1018XE
Amsterdam, The Netherlands}

\date{\today}

\begin{abstract}
After years of research into colossal magnetoresistant (CMR)
manganites using bulk techniques, there has been a recent upsurge
in experiments directly probing the electronic states at or near
the surface of the bilayer CMR materials
La$_{2-2x}$Sr$_{1+2x}$Mn$_2$O$_7$ using angle-resolved
photoemission or scanning probe microscopy. Here we report new,
temperature dependent, angle resolved photoemission data from
single crystals with a doping level of $x=0.36$. The first
important result is that there is no sign of a pseudogap in the
charge channel of this material for temperatures below the Curie
temperature $T_C$. The data show unprecedented sharp spectral
features, enabling the unambiguous identification of clear,
resolution-limited quasiparticle features from the bilayer split
3d$_{x^{2}-y^{2}}$-derived Fermi surfaces both at the zone face
and zone diagonal $k_F$ locations. The data show that these low
temperature Fermi surfaces describe closed shapes in $k_{||}$,
centered at the ($\pi/a$,$\pi/a$) points in the 2D Brillouin zone,
and are not open and arc-like in nature. The second important
result concerns the temperature dependence of the electronic
states. The spectra display strong incoherent intensity at high
binding energies and a very strong temperature dependence, both
characteristics reminiscent of polaronic systems. However, the
clear and strong quasiparticle peaks at low temperatures are
difficult to place within a polaronic scenario. Careful analysis
of the temperature dependent changes in the Fermi surface spectra
both at the zone face and zone diagonal regions in $k$-space
indicate that the coherent quasiparticle weight disappears for
temperatures significantly above $T_C$, and that the
$k$-dependence of the T-induced changes in the spectra invalidate
an interpretation of these data in terms of the superposition of a
`universal' metallic spectrum and an insulating spectrum whose
relative weight changes with temperature. In this sense, our data
are not compatible with a phase separation scenario.
\end{abstract}


\maketitle

The bilayered, strontium doped manganites
(La$_{2-2x}$Sr$_{1+2x}$Mn$_2$O$_7$) with $x\approx0.3-0.4$
(abbreviated forthwith LSMO) show on cooling an insulator-metal
transition associated with the onset of long range ferromagnetic
order. This transition occurs at a maximum Curie temperature
$T_{C}$, of approximately 130~K (for $x=0.36$) \cite{Ling, Kubota}
and goes paired with colossal changes in the magnetoresistance
(CMR) \cite{Moritomo}, which in turn have been linked to the large
number of spin and orbital degrees of freedom accessible to the
near-Fermi-level electronic states. Despite years of research, the
microscopic origin of the colossal magnetoresistance effect in
these systems is still the subject of much debate.

In general, metallic electrical transport requires the existence
of a Fermi surface (FS) upon which quasiparticles (QP's) reside.
In this context, LSMO is remarkable, as recent ARPES experiments
deep in the ferromagnetic, metallic state indicate the existence
of two highly differing regimes. Firstly, for $x=0.40$,
\cite{Shen} LSMO is reported to possess a discontinuous FS in the
form of arcs centered upon the Brillouin zone diagonal (ZD), with
a strong pseudogap opening up towards the Brillouin zone face
(ZF). In contrast, at an only 2-4\% lower doping level of
$x=0.36-0.38$, \cite{Dessau} the published spectra exhibit clear
quasiparticles at the ZF, leaving open the issue as to the
situation at the ZD.

A second, equally important issue regards the temperature
dependence of the QP's. Seeing as the metallic behaviour in LSMO
is symbiotic with the ferromagnetic order, one would expect
coherent QP's at the FS of LSMO to disappear at $T_C$. This
appears to be the case at the ZD in crystals with $x=0.40$.
\cite{Shen} Therefore, a recent report proposing both that ZF-QP's
exist well above $T_C$ for $x=0.36$ \cite{Dessau_NaturePhys} and
that metallic behaviour is seen in the ARPES spectra up to a $T$*
of 300K has stimulated much discussion.

The layered managanites play a special role within the CMR family,
as they enable exploitation of powerful, direct experimental
probes of the electronic states in both real (STM/STS) and
reciprocal space (ARPES). Therefore, it is of paramount importance
that the two central issues sketched above of whether the FS of
LSMO is pseudogapped below $T_C$, and whether QP's exist above
$T_C$ are investigated in detail.

This paper presents a thorough ARPES study of these key issues in
LSMO single crystals with $x=0.36$. Improved sample quality means
we are now able to unambiguously show the existence of QP's at the
zone face for both the antibonding and bonding c-axis bilayer
split 3d$_{x^{2}-y^{2}}$ bands and crucially, also for the ZD
direction in the same samples, meaning the
3d$_{x^{2}-y^{2}}$-derived FS has no pseudogap at low
temperatures. Furthermore, careful analysis of the $k$-dependence
of the data recorded at different temperatures provides strong
evidence against a scenario of phase separation into microscopic
metallic and insulating `patches' above $T_C$.
\cite{Dessau_NaturePhys}

Experiments used the UE112-PGMa beamline at BESSY coupled to an
SES100 analyzer and the SIS beamline at the Swiss Light Source
equipped with an SES2002 analyser and a 50 $\mu$m-sized light spot
($\mu$-ARPES). The total experimental energy broadening at 25K was
30~meV at BESSY and 20 meV at the SLS. The momentum resolution was
0.01~$\pi/a$ at the excitation energies used. Single crystals of
LSMO were grown in Amsterdam using the optical floating zone
technique. The Curie temperature of the samples was determined
using SQUID magnetometry to be 131~K. Prior to measurement, the
crystals were cleaved at $T<40$~K in a base pressure of
$1\times10^{-10}$~mbar. Very sharp, tetragonal low energy electron
diffraction patterns were obtained from all the measured cleavage
surfaces.

\begin{figure} [!tb]
\begin{center}
\includegraphics[width=1.0\columnwidth]{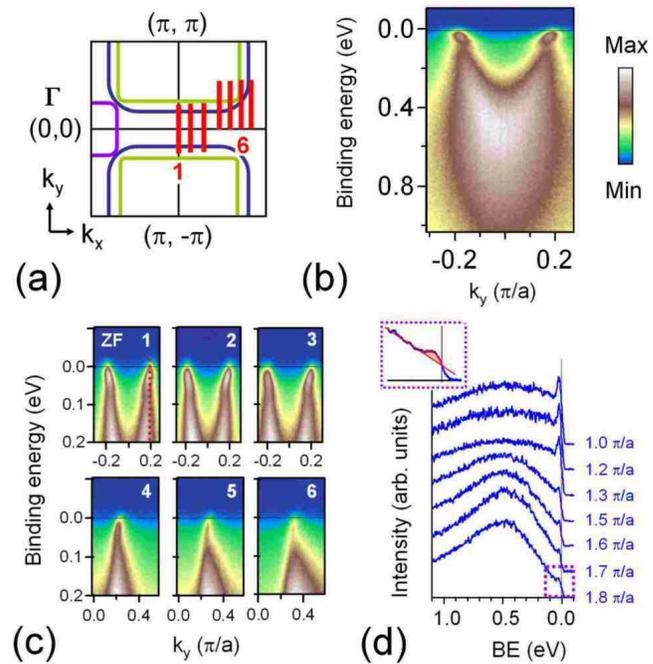}
\caption{\label{fig:EDM}(\textbf{a}) Schematic of the FS features
of LSMO, showing the 3d$_{z^{2}-r^{2}}$-based $\Gamma$-centered
pocket (purple) and the zone corner-centered FS barrels from the
3d$_{x^{2}-y^{2}}$-based bonding (green) and anti-bonding (blue)
bilayer split bands. The k-space cuts taken are marked in red.
(\textbf{b}) ($E,k$)-image from LSMO with $x=0.36$ taken at the
$(\pi,0)$ point in $k$-space [cut 1 in panel (a)]. (\textbf{c})
zoomed images taken at the $k$-slices shown in panel (a).
(\textbf{d}) $k_{F}$ EDCs from the images shown in (c). The inset
(highlighted in purple) shows a zoom of the EDC for
$k_{x}=1.8$~$\pi/a$. The red shading emphasizes that this EDC also
supports a small QP peak on top of a sloping background. All data
is recorded with $h\nu=56$~eV and $T=30$~K.}
\end{center}
\end{figure}

A typical ($E,k$)-image, taken at cut 1 of Fig. \ref{fig:EDM}a, is
shown in Fig. \ref{fig:EDM}b. A `U-shaped' band - in this case the
3d$_{x^{2}-y^{2}}$ anti-bonding (AB) bilayer-split band -
disperses from a band-bottom located at ca. $\approx450$~meV
towards the Fermi level, where each branch displays a sharp
feature at low energies indicative of the existence of QP's. In
panels (c) and (d) the $(E,k)$-images and energy distribution
curves (EDC's) from $k_{F}$ locations on the FS are shown. In the
former, it is evident that low energy spectral weight survives all
the way from the ZF to the ZD region of $k$-space.
\cite{AB_BB_footnote} In panel (d), the QP's show up as a small,
yet clearly discernable peak at $E_{F}$, followed by a hump, at
least part of which is due to emission from the deeper lying
bonding (B) band (see Fig. 2 and Ref. [\onlinecite{huang}]).

If one takes the presence of low energy spectral weight plus a
peaked EDC as a working definition of a QP-signal, it is evident
from Fig. 1 that the d$_{x^{2}-y^{2}}$ bands support QP's all
round the FS. Consequently, for $x=0.36$, these FS's are not
zone-diagonal `Fermi arcs' and do not support a pseudogap for
$T<T_C$. This is quite unlike the situation reported for
$x=0.40$,\cite {Shen} thus making pseudogap behavior for this
latter doping level not typical for the layered managanites in
general.

On a quantitative level, the clear reduction of the QP peak
intensity on going from the ZF to the ZD is intriguing.
\cite{nodal_comment} Future studies will be required to explore
whether these data can be taken as evidence of coupling to orbiton
or phonon-orbiton degrees of freedom, \cite{vandenBrinkPRL} which
would be expected to be maximal near the zone diagonal where the
3d$_{z^{2}-r^{2}}$ and 3d$_{x^{2}-y^{2}}$-derived FSs approach
closest to one another. \cite{huang}

In the preceding, two criteria were applied to the data to define
whether a QP existed or not. Naturally, it is of interest to
examine just how well defined these excitations appear to be in
both $k$ and $E$. It is known that effective cleavage surface
flatness - over the entirety of the illuminated area - can limit
the sharpness of ARPES features. The bilayer managanites do not
possess a Van der Waals bonded cleavage plane as the ARPES and STM
`standard' oxide Bi$_2$Sr$_2$CaCu$_2$O$_8$ does, and therefore
measurements on LSMO utilizing a highly focused `micro' excitation
spot could deliver better defined QP features.

\begin{figure} [!tb]
\begin{center}
\includegraphics[width=1.0\columnwidth]{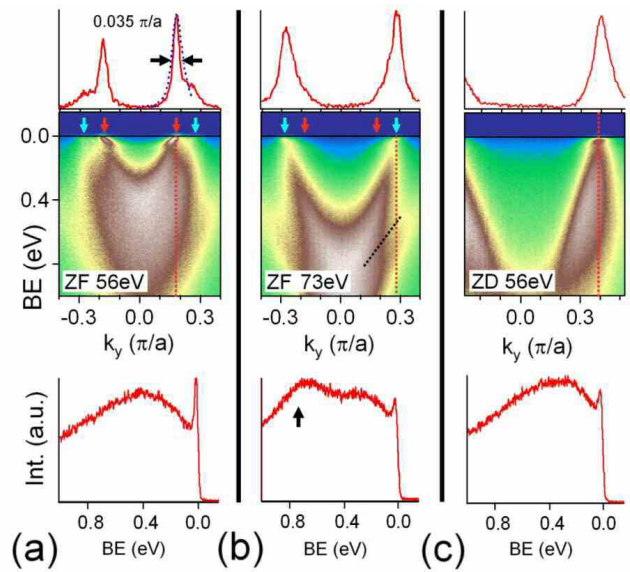}
\caption{\label{fig:kdep} $\mu$-ARPES data for LSMO with $x=0.36$
at $T=30$~K (\textbf{a}) at the zone face with $h\nu=56$~eV,
(\textbf{b}) $h\nu=73$~eV and (\textbf{c}) near the zone diagonal,
$h\nu=56$~eV. In each case, the upper panels contain $E_F$ MDCs
[in panel (a) overlaid with the $E_F$-MDC from the data shown in
Fig. \ref{fig:EDM}b in blue], the central panels the
$(E,k)$-images and lower panels the $k_F$ EDCs. The blue and red
arrows mark the $k_F$-location for the B and AB-bands,
respectively.}
\end{center}
\end{figure}

Figure \ref{fig:kdep} shows ($E,k$)-images recorded using the
50$\mu$m light spot at the SLS, together with their corresponding
$E_F$-MDC's (upper panels) and $k_F$-EDC's (lower panels). It is
clear from the ($E,k$)-image in (a) that the qualitative features
are very similar to those of Fig. \ref{fig:EDM}b. The big
advantage of the `$\mu$-ARPES' approach can be seen clearly in the
top panel of Fig. \ref{fig:kdep}a. Here the $E_F$ MDC's at
$(\pi,0)$ for the conventional and micro-spot measurements are
overlaid. The FWHM of the leading peak of the former is of order
0.06 $\pi/a$, a value equal to the state-of-the-art in the
literature (Refs. [\onlinecite{Dessau, Shen}]). The small spot
data is almost two times narrower, with a FWHM of only 0.035
$\pi/a$. The corresponding $k_F$ EDC shows a more pronounced peak
at $E_F$, whose width is essentially resolution-limited, as would
be expected for a quasiparticle.

Fig. \ref{fig:kdep}b shows the same $(\pi,0)$ cut for
$h\nu=73$~eV. This photon energy is reported to strongly favour
emission from the B band, \cite{Dessau} and indeed a further
U-like feature is seen, with a band bottom located at ca. 800~meV
and $k_F$-wavevectors some 0.1 $\pi/a$ greater than those for the
AB band. The B band $k_F$ EDC shows a small, narrow peak at $E_F$
signalling the QP, followed by two hump-like features at higher
energy.

Consideration of both panel (a) and (b) of Fig. \ref{fig:kdep},
reveals a surprising feature of the $\mu$-ARPES measurements on
LSMO in the form of an extra, V-shaped, weak feature indicated in
the ($E,k$)-image of Fig. \ref{fig:kdep}b with a black dotted
line. This feature disperses in a parabolic manner around the
$(\pi,0)$-point (as do the AB and B bands). From an extrapolation,
its $k_F$ exceeds 0.5~$\pi/a$, a value that is far from matching
any FS crossing in band structure calculations. \cite{huang} At
present the origin of this feature, which gives rise to the high
energy hump in the EDC indicated with an arrow in Fig.
\ref{fig:kdep}b, is unknown. Both the simplicity of the observed
tetragonal LEED patterns for the cleaved surfaces and the k-space
location argue against a simple `diffraction replica' origin.

Fig. \ref{fig:kdep}c shows data taken in the region of the ZD with
$h\nu=56$~eV. Once more a sharp QP peak is seen at $E_F$, whose
intensity, although greater than in the conventional measurement
shown in Fig. \ref{fig:EDM}c and d, is still small compared to
that seen at the ZF for the same photon energy.

Given the identification of well-defined QP features for both
$(\pi,0)$ centered FSs, one can then move on to examine their
nesting characteristics. Fig. \ref{fig:EDM}c shows significant
nesting in the ZF region for the AB band (the $k_F$ vectors remain
constant over a fair range in $k_x$). The nesting vector
(expressed in units of $2\pi/a$) is simply the $k_F$ value of
$q=0.19$. For the B band at ($\pi$,0), the analogous value would
be $q=0.28$, making it a much closer match with q-vectors at which
Jahn-Teller correlations above $T_C$ give broad regions of
incommensurate x-ray scattering centered at ($\pm\epsilon, 0,
\pm1)$ with $\epsilon=0.3$. \cite{Campbell} This may lie behind
the significantly weaker intensity of the QP feature for the B
band EDC as compared to the analogous AB band EDC seen in Fig.
\ref{fig:kdep}, although one should bear in mind that the polaron
correlations seen in x-ray experiments collapse for $T<T_C$.
\cite{Campbell}

To summarize this part of the results, the observation of
persistent and resolution-limited QP signals at all $k_F$
locations probed on the 3d$_{x^{2}-y^{2}}$-derived FSs using
$\mu$-ARPES measurements is not only fully consistent with
non-gapped $(\pi,0)$-centered FSs in LSMO with $x=0.36$ at low
temperature, but also sets the standard for the sharpest
photoemission data on the bilayer manganites recorded to date.

\begin{figure} [!t]
\begin{center}
\includegraphics[width=1.0\columnwidth]{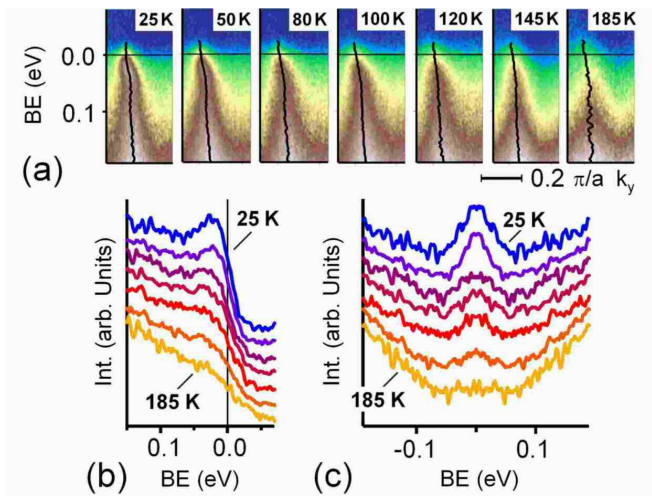}
\caption{\label{fig:Tdependence_pic4}(\textbf{a}) T-dependence for
LSMO $x=0.36$ at ($\pi$,0) (\emph{h$\nu$=$56$}~eV, left hand
branch). The MDC peak maxima are overlaid using a black solid
line. (\textbf{b}) EDCs at $k_{F}$ for all temperatures,
normalized to the intensity at 150~meV BE, and (\textbf{c}) their
symmetrized versions (all EDCs are offset vertically). }
\end{center}
\end{figure}

We now turn our attention to the issue of the temperature
dependence of the Fermi surface electronic states.
Fig.~\ref{fig:Tdependence_pic4} shows data from the
$(\pi,0)$-point taken at temperatures from 25~K up to 185~K, the
latter well above the $T_{C}$ of 131~K. Panel (a) contains EDMs
showing one of the branches of the AB band. It is clear that --
apart from thermal broadening -- all the EDMs look comparable. It
should be noted that the spectral weight of the QP peak does
decrease steadily with temperature, as can be seen from the EDCs
shown in Fig.~\ref{fig:Tdependence_pic4}b. We note here that this
decrease in intensity with higher temperatures is larger than the
change in thermal broadening can account for on going from 30 to
185~K. Nevertheless, even at 185~K some spectral weight at $E_F$
still remains as a step in the EDC. None of the spectra in
Fig.~\ref{fig:Tdependence_pic4} show a gap within the experimental
resolution and the accuracy of the energy referencing used. This
can be seen from the symmetrized $k_{F}$-EDCs displayed in panel
(b), although the peak in the 185~K symmetrized spectrum is very
small and this EDC might actually be close to being gapped,
hinting that a true gap in the charge sector might be opening at
even higher temperatures. The fact that spectral intensity at
$E_F$ near $(\pi,0)$ exists even 50~K into the paramagnetic region
of the phase diagram\cite{Dessau_NaturePhys} seems at odds with
the globally insulating transport characteristic at these elevated
temperatures of bilayered LSMO with $x=0.36$. Therefore, drawing
together what our T-dependent data show us up to this point: (i)
the QP-peak at lowest energy disappears at or very close to the
bulk Curie temperature, yet (ii) at the zone face, there remains
(non-peaked) spectral weight at $E_F$ until temperatures of the
order of 50K above $T_C$.

\begin{figure} [!b]
\begin{center}
\includegraphics[width=1.0\columnwidth]{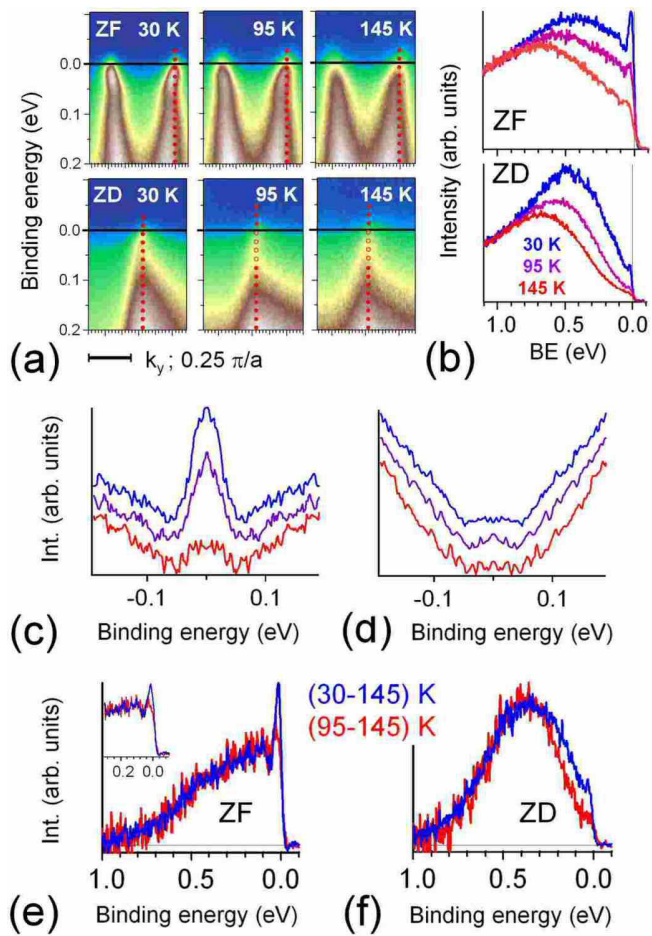}
\caption{\label{fig:Tdep} (\textbf{a}) EDMs of LSMO ($h\nu=56$~eV,
$k_{x}=1.0\pi/a$) at the ZF and ZD point in $k$-space at three
different temperatures. (\textbf{b}) EDCs at $k_{F}$ from the EDMs
shown in panel (a) (indicated there with red dotted lines). EDCs
are normalized to the high BE intensity, which is equivalent to
normalization to the intensity above $E_F$. (\textbf{c})
Symmetrized EDCs near $E_F$ (scaled at intensity at 200~meV BE)
for the three temperatures at the ZF and (\textbf{d}) at the ZD.
(\textbf{e}) The scaled difference between the 30 and 95~K EDC
(blue curve) and the 95 and 145~K EDC (red curve) for the ZF. The
inset shows the same spectra over 300~meV BE for EDCs that are
first broadened to match the Fermi-Dirac-distribution at 145~K
before subtraction (see text). (\textbf{f}) The scaled difference
EDCs of $30-145$~K EDC (blue curve) and $95-145$~K EDC (red curve)
for the ZD.}
\end{center}
\end{figure}

To further investigate this anomalous temperature dependence,
spectra were taken with improved statistics and over a wider
binding energy range, both the ZF \emph{and} ZD regions of the
$(\pi,\pi)$-centered Fermi surfaces at low temperature and just
below and above $T_C$. These data are shown in Figure
\ref{fig:Tdep}. Panel (a) contains the (zoomed) EDMs and (b) the
accompanying $k_{F}$ EDCs.

In panels (c) and (d) of Fig. \ref{fig:Tdep} symmetrized versions
are depicted of the EDCs shown in panel (b). Although, as asserted
below, the low binding energy coherent spectral weight disappears
steadily with increasing temperature at both $k$-points, the
symmetrized spectra show that at the ZF and the ZD again no gap in
the charge sector is opening at temperatures up to 140~K.

One striking feature in these T-dependent data, both at the ZF and
ZD regions, is that the aforementioned reduction in spectral
weight extends over an energy scale of up to 700-800~meV, which is
obviously an energy range far in excess of the change in thermal
energy. Such large changes have been reported previously for the
ZF region in Ref. [\onlinecite{Dessau_NaturePhys}], and their
T-dependence has been put forward as an argument for microscopic
phase separation, whereby local metallic and insulating regions
change their relative abundance steadily with temperature. Vital
for this model is that the difference spectra between pairs of
$k_F$ EDCs (recorded at different temperatures) can be collapsed
onto a single, universal `metallic EDC'. \cite{Dessau_NaturePhys}
The two difference spectra from our data recorded from the ZF
region are shown in Fig. \ref{fig:Tdep}e. At first glance, they do
look to be scaled versions of one another, but closer inspection
reveals that the weight of the QP-peak in the original EDCs
decreases much faster between 95 and 145~K than it did between 30
and 145~K, meaning that the difference spectra lose their low
energy peak as temperature increases. As illustrated in the inset
of Fig. \ref{fig:Tdep}e, this decrease in spectral weight of the
QP in the difference spectra persists -though less prominently- if
the original EDCs taken at 30 and 95~K are first broadened using
an additional Fermi-Dirac distribution (matching the temperature
broadening of the spectra recorded at 145~K) before the
subtraction takes place. This proves that the decrease in QP
spectral weight cannot be attributed to temperature broadening
alone. If, for the sake of argument, one were to assign the
difference spectra in Fig. \ref{fig:Tdep}e to a metallic phase as
was done in Ref. [\onlinecite{Dessau_NaturePhys}], the data here
show clearly that this `metallic' spectrum loses its coherent QP
spectral weight as the temperature climbs above $T_C$.

An additional, powerful test of this phase separation scenario is
given by the $k$-dependence of the spectral changes with
temperature. If temperature essentially only alters the relative
percentage of the metallic and insulating patches, then various
difference spectra at a FS location other than the ZF should also
collapse onto a universal `metallic EDC'. From Figs. \ref{fig:EDM}
and \ref{fig:kdep}, it is clear that LSMO with $x=0.36$ also
supports QPs in the zone diagonal region, thus offering the chance
of an independent test of the phase separation scenario.

The ZD difference spectra shown in Fig. \ref{fig:Tdep}f clearly
differ significantly up to binding energies of 300~meV, thus one
cannot speak of a universal `metallic EDC' growing monotonically
with decreasing temperature for the ZD region of $k$-space. This,
in turn, provides a strong argument against the phase separation
model presented in Ref. [\onlinecite{Dessau_NaturePhys}] as being
a complete description of the T-dependence of these $k$-dependent
data, as this scenario would demand that the change in spectral
weight happens in a similar fashion at both high symmetry points
of the Brillouin zone. At this stage, it is interesting to recall
that a recent T-dependent STM study of LSMO with $x=0.30$
\cite{Renner} found no signs of a bimodal gap distribution in
tunnelling data across more than 2000 different surface locations,
which - taken at face value - also argues against phase
separation.

Thus, the challenge faced is to reconcile the following four main
characteristics of these new, highly-resolved LSMO data for
$x=0.36$:

\renewcommand{\theenumi}{\roman{enumi}}

\begin{enumerate}
\item Existence of clear and sharp QP peaks at all
3d$_{x^{2}-y^{2}}$-derived $k_F$ locations probed at low T.
\item Strong high BE spectral weight at all temperatures.
\item T-dependent changes in the spectra on
a scale of up to 800~meV BE.
\item Clear differences in the
temperature induced effects between the ZF and ZD regions of
$k$-space.
\end{enumerate}

Phase separation as described above would have offered a route to
reconcile points (i)-(iii), but cannot account for point (iv). On
the other hand, behavior such as described in points (ii)-(iv) has
been reported for photoemission from polaronic systems,
\cite{Kim,KShen} in which most of the higher binding energy
intensity is due to multiphonon satellites, and a vanishingly weak
QP (the `zero phonon line') is left at $E_F$. The kind of
electron-phonon coupling strengths required to generate all the
observed high BE incoherent weight \cite{footnote_hiBE} seen in
these data from LSMO would lead to the complete suppression of the
QP feature at $E_F$, something that evidently does not occur here
for $T<<T_C$ [point (i), above].

From the above, it is apparent that some pieces of the puzzle that
are required for a complete and consistent description of the
ARPES data from these systems (and their temperature dependence)
must still be missing. Despite this pointer towards future work in
this area, the data do result in a number of unambiguous and
important conclusions. Firstly, that the $(\pi,\pi)$-centered
Fermi surfaces of the layered CMR manganite
La$_{1.28}$Sr$_{1.72}$Mn$_2$O$_7$ support quasiparticles, both at
the diagonal and the face of the 2D Brillouin zone at low
temperatures. This system, therefore, has no pseudogap in the
charge sector, thereby excluding the use of the epithet `nodal
metal' for this class of materials in general. Secondly, we find
that the temperature dependent behavior of both the quasiparticles
and of higher lying spectral weight is different at the Brillouin
zone face and diagonal. This argues against a model describing the
metal-insulator transition as stemming from a percolative growth
of metallic clusters in an insulating matrix.

\par\ \par

Our thanks to the IFW Dresden group for lending us their
spectrometer, and to W. Koops and T. J. Gortenmulder for expert
technical support. This work is funded by the FOM (ILP and SICM)
and the EU (via I3 contract RII3-CT-2004-506008 at both BESSY and
SLS).


\begin{thebibliography}{}
\bibitem{AB_BB_footnote} For the ZD direction, the bonding and
antibonding bilayer split FS's almost touch. Exactly which of
these FS's dominates the observed intensity is a moot point.
\bibitem{nodal_comment} We note here that a reduction of intensity
at the ZD in the ARPES experiment could also be (at least
partially) the result of: (i) matrix element effects (ii) the fact
that the $k$-space cut at the ZD is not perpendicular to the FS
and (iii) the increasing Fermi velocity towards the ZD region of
the Brillouin zone.
\bibitem{footnote_hiBE}
We note that the $k_F$ EDCs as presented in this paper also
contain spectral weight due to bands located at higher BE, for
example the B band and the third, at present unidentified feature
mentioned earlier.
\end{thebibliography}
\end{document}